\documentclass[preprint,12pt]{elsarticle}

\usepackage{amssymb}
\usepackage{amsmath}
\usepackage{listings}
\usepackage{xcolor}
\lstdefinestyle{pythonstyle}{
    backgroundcolor=\color{white}, 
    basicstyle=\footnotesize\ttfamily,
    breaklines=true,
    keywordstyle=\color{blue},
    commentstyle=\color{green},
    stringstyle=\color{red},
    morekeywords={ALL_TRIPLES_KNOWLEDGE_BASE, ADD, IN, TO, EMPTY_LIST, REMOVE, FROM, FOR, IF, ELIF, RETURN, END, MAP, remove, and, not in, Remove, from},
    literate={∪}{{\ensuremath{\cup}}}1
             {∅}{{\ensuremath{\emptyset}}}1
}
\usepackage{algorithm}
\usepackage{algpseudocode}
\usepackage{amsmath}

\journal{Computers \& Mathematics with Applications}
\usepackage{pdflscape}  
\usepackage{soul}
\sethlcolor{white}
\usepackage[most]{tcolorbox}

\newtcolorbox{highlightbox}{
  colback=yellow!20,
  colframe=yellow!20, 
  boxrule=0pt,        
  arc=0pt,            
  left=2pt, right=2pt, top=2pt, bottom=2pt,
}
\begin{document}

\begin{frontmatter}



\title{Polynomiogram: An Integrated Framework for Root Visualization and Generative Art} 

\author[1]{Hoang Duc Nguyen}
\author[2]{Anh Van Pham}
\author[3,4]{Hien D. Nguyen\corref{mycorrespondingauthor}}
\cortext[mycorrespondingauthor]{Corresponding author: Hien D. Nguyen (hiennd@uit.edu.vn)}

\affiliation[1]{organization={Georgia Institute of Technology},
             city={Atlanta},
             postcode={30332},
             state={Georgia},
             country={USA}}

\affiliation[2]{organization={ECO Vietnam Group},
             city={Ho Chi Minh city},
             postcode={700000},
             country={Vietnam}}

\affiliation[3]{organization={University of Information Technology},
             city={Ho Chi Minh city},
             postcode={700000},
             country={Vietnam}}

\affiliation[4]{
    organization={Vietnam National University},
             city={Ho Chi Minh city},
             postcode={700000},
             country={Vietnam}}

\begin{abstract}
This work introduces the Polynomiogram framework, a unified computational system for exploring polynomial root structures and generating algorithmic art. Its novelty lies in a flexible two-parameter sampling scheme, where parameters drawn from user-defined domains mapping to polynomial coefficients via a generating function. This approach bridges scientific analysis and creative visualization within the same mathematical foundation. The framework combines NumPy’s companion-matrix solver for large-scale efficiency and MPSolve for high-precision validation, ensuring both computational speed and numerical analyzing. Beside analysis, the Polynomiogram produced personalized generative artworks, such as natural forms and AI-inspired compositions, showcasing its capacity to merge symbolic modeling, root computation, and visual creativity in a single, versatile platform.
\end{abstract}

\begin{graphicalabstract}
\includegraphics[width=\textwidth]{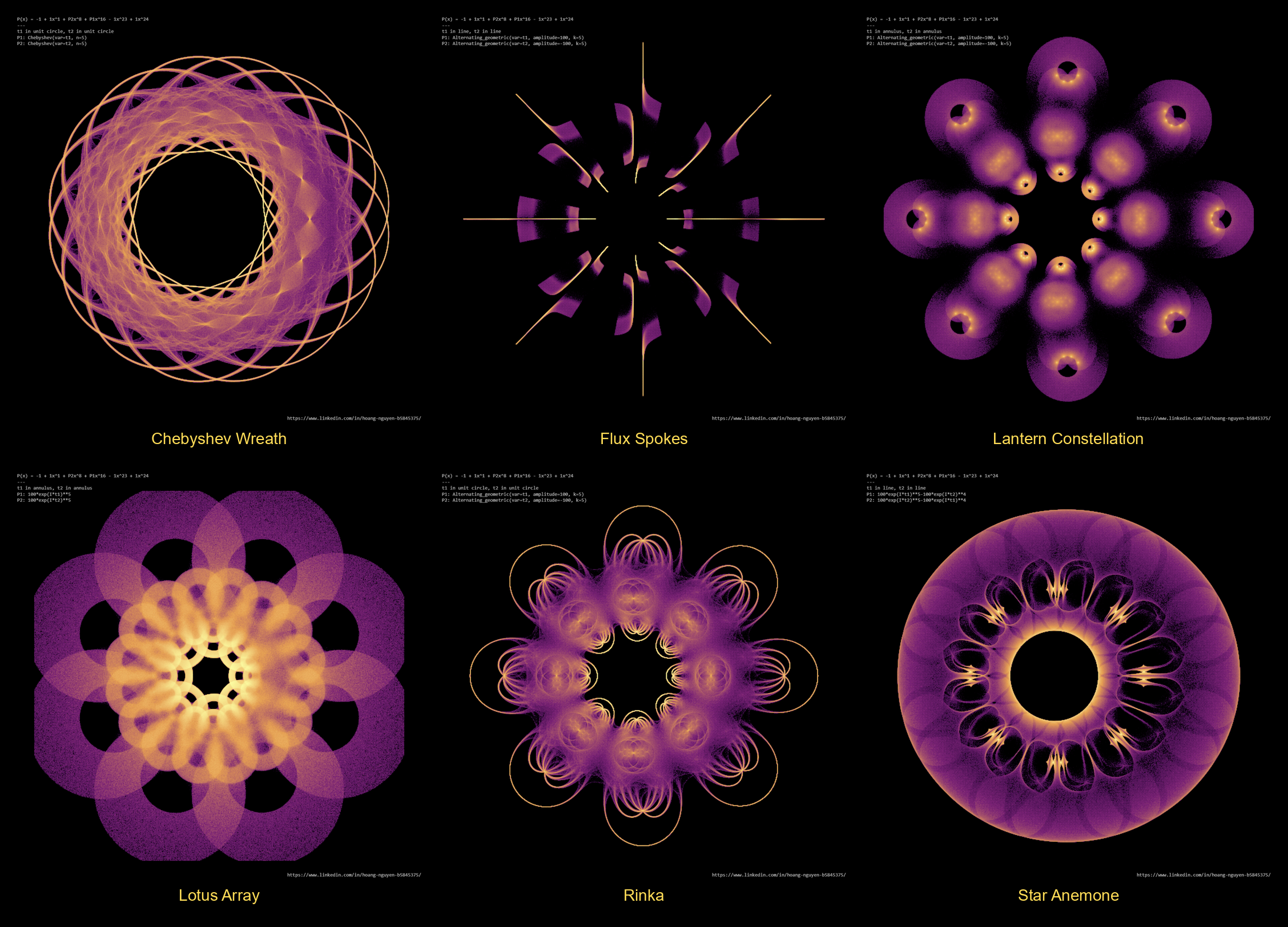}
\end{graphicalabstract}

\begin{highlights}
\item New mathematical visualization object, Polynomiogram, which maps the global distribution of polynomial roots over parameterized families.
\item The framework is a full-stack computational system that links symbolic to numeric pipeline.  
\item Dual numerical engines for accuracy and scale.
\item Demonstrated capability for Generative Art.
\end{highlights}

\begin{keyword}
Soft Computing \sep Computer Algebra \sep Multi Precision Computation  \sep Symbolic to Numeric Pipeline  \sep Polynomiogram \sep Generative Art. 



\end{keyword}

\end{frontmatter}

\section{Introduction}
\label{sec_intro}
In many scientific and engineering systems, parameters determine whether the behavior is stable, oscillatory, or chaotic. Bifurcation theory provides the mathematical foundation for studying such transitions by analyzing how characteristic parameters influence system equilibria and their stability \cite{kuznetsov1998elements}. Although this theory is a mature and exciting topic, its computational counterpart, continuation and curve tracing, remains largely confined to academic study. These methods deliver precise local information about individual branches but are difficult to scale or parallelize when parameters vary widely.

In parallel, another long-standing line of research has investigated the global organization of polynomial roots. Rather than following how individual roots move with parameters, these studies examine how roots are distributed collectively across entire families of polynomials. Classical examples include random polynomials, where coefficients are treated as random variables, and lacunary or sparse polynomials, where many terms disappear \cite{kac1943average,schinzel1970reducibility}. Theoretical work by Kac, Edelman, and Kostlan showed that random polynomials with Gaussian coefficients tend to cluster in the unit circle, whereas lacunary polynomials exhibit distinctive root gaps and reduced numbers of real roots. Orthogonal and recursive polynomial systems, such as Chebyshev and Lucas families, form another major branch of this research \cite{mason2002chebyshev,nalli2009generalized}. Together, these traditions established a statistical and geometric understanding of root distributions but remained largely analytical, relying on probabilistic models and asymptotic reasoning rather than
visualization.

At the opposite end of the spectrum lies digital art, where mathematical formulas have long been used to create images \cite{briggs1992fractals}. Fractals, strange attractors, and algorithmic geometry illustrate how simple deterministic rules can produce intricate visual structures. Polynomiography, introduced by Bahman Kalantari in the early 2000s, turned iterative root-finding algorithms such as Newton’s method into visual art by mapping their basins of attraction \cite{kalantari2005polynomiography}. It revealed the aesthetic potential of mathematical computation but remained centered on the dynamics of iteration. 

The present work extends this idea in a new direction and introduces the concept of the
Polynomiogram. The name combines polynomial with the suffix -gram, from the Greek gramma, meaning “something written or drawn.” The same suffix appears in words such as spectrogram and phase diagram, which denote visual representations of mathematical or physical structures. Whereas Polynomiography visualized the trajectories of iterative solvers, a Polynomiogram visualizes the global distribution of polynomial roots across parameterized families. It replaces the local dynamics of iteration with a statistical sampling of parameters and constructs density maps that represent how the roots populate the complex plane. These root-density fields reveal geometric and spectral structures that can be interpreted scientifically as indicators of stability or aesthetically as algorithmic patterns. In this sense, the Polynomiogram serves as a bridge between the analytical tradition of bifurcation and the creative exploration of digital art.

The key contributions of this work are threefold. First, it introduces a general statistical methodology for visualizing polynomial spectra through global root-density maps, enabling a quantitative and aesthetic interpretation of root distributions. Second, it develops a comprehensive computational framework that seamlessly integrates symbolic model generation, parallel eigenvalue solving, and high-resolution visualization, with optional multi-precision validation for scientific rigor. Finally, it demonstrates the versatility of the Polynomiogram across domains from stability analysis in engineering systems to algorithmic and generative art, which highlights its role as a unifying bridge between mathematical analysis and creative expression.

The next section reviews related work in polynomiography, spectral visualization, and root-finding algorithms. Section 3 describes the framework, including symbolic representation, solvers, sampling strategies, and visualization methods. Section 4 presents results, discussion, and applications. The last section concludes this study and discuss future improvements.

\section{Background}

Polynomial root finding is one of the oldest problems in mathematics and remains a central topic in computational science. The task of determining the zeros of a polynomial arises in fields as diverse as physics, chemistry, engineering, and among others. Early efforts focused on deriving exact algebraic formulas, while later work turned toward numerical, iterative, and probabilistic strategies as the complexity of higher degree equations became clear. The following sections trace this progression, beginning with analytic formulas and continuing through modern computational techniques that combine linear algebra, numerical refinement, and symbolic computation.

Early mathematicians developed explicit formulas for quadratic, cubic, and quartic equations, expressing the roots in terms of radicals. These achievements, refined through the works of Cardano and Ferrari in the sixteenth century, marked the peak of classical algebraic theory. However, the search for a general formula for higher degrees eventually reached a barrier. The Abel- Ruffini theorem demonstrated that no general radical solution exists for polynomials of degree five or greater, establishing a fundamental limit of algebraic solvability \cite{trefethen2022numerical}. This result redirected mathematical interest from symbolic formulas to numerical computation, where approximation and algorithmic design replaced exact manipulation.

The recognition that higher degree equations cannot be solved by radicals shifted attention from symbolic manipulation to numerical computation. Mathematicians began to design procedures that approximate roots through algebraic transformations or repeated refinement. Two main directions emerged. The first treats root finding as an eigenvalue problem by representing the polynomial through a matrix whose eigenvalues correspond to its roots. The second relies on iterative updates that refine approximate roots through repeated correction. These two strategies provided the foundation for advances in computational algebra.

The first approach represents the polynomial through the Frobenius companion matrix, a dense square matrix whose structure reflects the polynomial coefficients. This formulation converts the root finding task into a matrix eigenvalue problem, opening the way to systematic analysis using linear algebra. In this framework, the polynomial roots are obtained from the characteristic equation of the matrix, and the numerical effort shifts to computing its eigenvalues. The stability and efficiency of this process were made practical by the introduction of the QR algorithm, a numerical procedure that repeatedly applies orthogonal transformations to reduce a matrix to upper triangular form while preserving its eigenvalues. A related approach, the Schur decomposition, expresses a matrix as the product of a unitary matrix, an upper  triangular matrix, and the inverse of the unitary matrix, providing a stable foundation for modern eigenvalue
solvers. These routines were later standardized and optimized in the Linear Algebra Package (LAPACK), which forms the computational core of many scientific libraries. The computational cost of this dense eigenvalue calculation scales approximately as $O(n^3)$ for a polynomial of degree $n$ \cite{edelman1995polynomial}. Structured variants of this idea, such as the colleague and Fiedler matrices, improve numerical conditioning when polynomials are expressed in orthogonal bases and exploit matrix sparsity to enhance stability \cite{good1961colleague,fiedler2003note}.

While the matrix formulation provided a unified and reliable way to compute all roots simultaneously, its computational cost and sensitivity to coefficient scaling limited its use for very large or poorly conditioned polynomials \cite{press2007numerical}. Another approach explored the direct iterative refinement, where each root or set of roots is improved through repeated numerical updates rather than through matrix factorization. These algorithms became the second major direction in numerical root finding and laid the groundwork for a variety of modern refinement and hybrid methods.

The earliest and most widely known example is Newton’s method, which refines one root at a
time by linearizing the polynomial near an initial guess. When good starting points are available,
this method converges rapidly, but its success depends heavily on the choice of initial values. To
make the process more reliable, hybrid schemes were introduced that combine iteration with
deflation, removing each converged root from the polynomial to reduce numerical interference
among remaining roots. The best known of these hybrid solvers is the Jenkins - Traub algorithm,
which divides the computation into three coordinated stages (iteration, quadratic convergence,
and deflation) and achieves consistent robustness across diverse coefficient scales \cite{jenkins1970three}.

As computers grew more powerful, attention turned toward methods that could refine all roots simultaneously instead of one at a time. The Durand - Kerner and Aberth - Ehrlich algorithms update every root through coupled iterations, allowing parallel computation and often faster overall convergence\cite{henrlci1977applied}. These simultaneous strategies remain appealing because of their conceptual simplicity and inherent parallelism.

Early iterative methods, designed to efficiently compute roots, gave rise to concepts like Polynomiography, introduced by Kalantari \cite{kalantari2005polynomiography}. This visualization technique maps the complex plane, assigning colors to points based on which polynomial root the iteration converges to. These resulting images clearly expose the geometry of basins of attraction and dramatically illustrate the method's sensitivity to initial conditions. This visual perspective connects numerical analysis with mathematical visualization, highlighting the dynamic nature of polynomial solving.

While iterative and hybrid methods improved speed and flexibility, another line of research emerged that aimed for mathematical certainty rather than numerical speed. For real polynomials, certified computation is achieved through subdivision and isolation techniques that gradually narrow the intervals containing the roots. These methods divide the real line into smaller segments and determine how many roots lie within each segment. Classical approaches include Sturm sequences, Descartes rule of signs, and continued fraction refinements, all of which count or bracket roots within specific intervals and then refine these intervals until every real root is isolated \cite{collins1976polynomial}. Closely related techniques extend this idea into the complex plane through interval arithmetic. The interval Newton and Krawczyk operators apply verified numerical analysis to enclose each complex root within a bounded region while ensuring that all roots are captured and none are duplicated \cite{moore1959interval}.

By the late twentieth century, as computer algebra systems matured, a new direction of research began to move beyond numerical representation. These methods relied on exact algebraic manipulation combined with controlled randomness to perform root finding symbolically. They used exact number systems such as rational arithmetic to eliminate the precision problems of floating-point computation. Randomization served as a practical tool for exploring solution spaces and carrying out algebraic operations such as factorization more efficiently, while still guaranteeing correctness once the algorithm completed. Within this framework, probabilistic algorithms such as the Las Vegas methods became central tools for symbolic computation \cite{babai1979monte}. These algorithms use random choices to simplify or accelerate complex algebraic tasks while guaranteeing exact results upon successful termination. Important examples include the work of Kaltofen, which applied randomization to polynomial factorization and exact algebraic solving \cite{kaltofen1995subquadratic}. Although these techniques are less common in standard numerical libraries, they demonstrate how randomness can be harnessed to accelerate and stabilize difficult algebraic computations.

Beyond static root finding, another area of research examines how the roots of a polynomial change as parameters vary. In classical bifurcation theory, many stability problems can be reduced to the study of characteristic polynomials that arise from the linearization of nonlinear systems. The locations and multiplicities of these roots determine whether equilibria are stable and when bifurcations occur. Root crossings along the real or imaginary axis indicate transitions such as the onset of oscillations or multiple steady states. Visualization of these root trajectories has long been used to understand qualitative changes in system behavior.

Applications in chemical and physical systems often make use of this connection indirectly. For example, researchers analyzed the multiplicity and sensitivity of steady states in reaction models, where the underlying stability conditions depend on the roots of characteristic polynomials \cite{moniruzzaman2025forced,sun2018bifurcation}. These studies demonstrate how the movement or coalescence of roots can signal changes in dynamic behavior. Besides bifurcation and continuation studies, another group of research traditions examines the global organization of polynomial zeros. Rather than following how individual roots move with parameters, this research school seeks to understand how entire sets of roots are distributed across families of polynomials. Although these studies share the same motivation of describing root organization and global patterns, they differ in mathematical structure and purpose \cite{kac1943average,edelman1995many,redei2014lacunary,koshy2019fibonacci}. Most studies have been analytical in nature, focusing on probabilistic models and asymptotic results.

The desire to see mathematics together with describing it has long driven the field of mathematical visualization. A landmark example is fractal geometry, which revealed that simple recursive formulas can generate images of extraordinary complexity. The Mandelbrot and Julia sets demonstrated how self-similarity and sensitivity to initial conditions can be expressed visually, linking dynamical systems and complex analysis to art. These discoveries transformed mathematics into something that could be seen as well as proved \cite{peitgen1986beauty}. This merging of computation and imagery eventually found a home in computer art. Algorithmic visualization tools such as Ultra Fractal, Apophysis, and Chaoscope allow users to explore iterative systems and create intricate fractal designs without requiring deep mathematical training. Although these programs have shaped digital art for decades, their focus lies in dynamical iteration rather than in the structure of algebraic equations. They do not include symbolic parameterization, large-scale numerical root solvers, or data-driven visualization of polynomial families.

The work presented here builds on this visual tradition but directs it back to the algebraic world. The Polynomiogram framework is designed specifically to study and visualize polynomial families. The following section describes the framework in detail, outlining its symbolic structure, numerical strategies, parallel sampling, and rendering process.

\section{Framework of the Polynomiogram system for large-scale visualization}
\label{sec:framework}

\begin{figure}[htbp]
    \centerline
    {\includegraphics[width=0.75\textwidth]{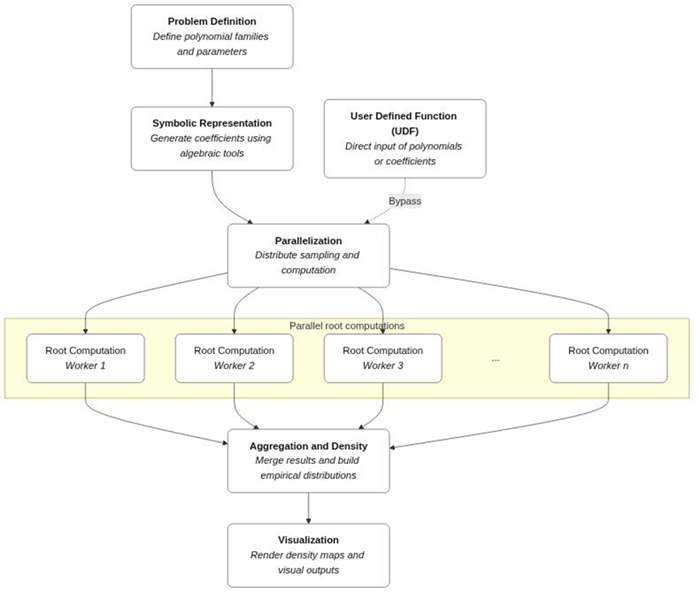}}
    \caption{Conceptual workflow of the Polynomiogram framework.}
    \label{figWorkflow}
\end{figure}

This section introduces the conceptual framework of the Polynomiogram system for large-scale visualization of polynomial roots. The framework operates as a complete pipeline that links algebraic definition, numerical computation, and visual analysis. It consists of six main stages described in Figure \ref{figWorkflow}: problem definition, symbolic representation, root computation, parallelization, aggregation and density mapping, and visualization.

In the standard workflow, polynomial families are defined symbolically, parameters are sampled within bounded domains, and all roots are computed and visualized as density maps in the complex plane. An optional user-defined function (UDF) interface allows users to bypass the symbolic and sampling stages by supplying explicit polynomials or coefficient sets directly, proceeding straight to root computation and visualization. Implementation details and reproducible code for the standard workflow are provided in the accompanying Github repository; the description here focuses on the conceptual design and data flow of the system.

Consider univariate complex polynomials of the form:
\begin{equation}
    p(x,\theta):=\sum_{k=0}^{n}a_k(\theta)x^k
\end{equation}
where, $n$ is the fixed degree and $a_k(\theta)$ are parameter dependent coefficients. Polynomials may be represented in dense or sparse form; coefficients with zero value are omitted implicitly. Samples for which the leading coefficient $a_n(\theta)$ vanishes are excluded to ensure that only well-defined equations are analyzed.

The parameters $\theta$ are drawn from bounded domains, and each coefficient function $a_k(\theta)$ is assumed continuous. These assumptions guarantee that all roots remain within a finite region of the complex plane, enabling stable numerical computation and meaningful visualization of their aggregated density.

To describe these parameters more concretely, the framework introduces two latent variables, $\theta=(t_1,t_2)$, which serve as the primary controls for generating polynomial coefficients. Each variable represents a complex value selected from a bounded region of the complex plane, and together they define a coordinate pair that specifies one polynomial instance within the family. For each sampled pair, $(t_1,t_2)$, the coefficient functions $a_k(\theta)$ are evaluated to produce a unique polynomial.

This formulation establishes a continuous relationship between the parameter space and the coefficients, allowing the polynomial structure to change smoothly as the parameters vary. Coefficients may be defined directly through algebraic expressions, such as $a_k=t_1^2+i*t_2$, or generated from analytical templates, including Chebyshev or alternating geometric forms. Both approaches preserve coherence among neighboring samples, ensuring that gradual changes in parameter values correspond to gradual transformations in root patterns.

Parameter sampling follows several well-defined geometric configurations in the complex plane. The unit circle introduces angular variation, the disk extends this to include radial effects, the annulus provides bounded radial control, and the line segment creates a simple linear path useful for studying boundary behavior. Sampling within these bounded regions is uniform to preserve statistical balance. Fixed pseudo-random seeds ensure reproducibility, and independent random generators for $t_1$ and $t_2$ prevent correlation between variables.

Building on this foundation, the framework extends the sampling strategy by allowing the two latent parameters $t_1$ and $t_2$ to be drawn either from the same domain or from different ones. When both variables share the same sampling region, the resulting parameter space exhibits uniform and symmetric geometric properties. The framework also supports heterogeneous configurations in which $t_1$ and $t_2$ originate from distinct domains. In such cases, the parameter space forms the Cartesian product $D_1 \times D_2$, where $D_1$ and $D_2$ represent the respective sampling regions. Each sampled pair $(t_1,t_2)$ defines one polynomial, and the entire collection of pairs constitutes the ensemble used for statistical and visual analysis.

Allowing distinct sampling domains for the two latent variables is a key and original contribution of this work. To the best of our knowledge, this is the first framework that systematically supports mixed domain parameterization for large-scale polynomial root visualization. Because the coefficients $a_k=f_k(t_1,t_2 )$ depend on both variables, different combinations of sampling domains generate independent geometric influences within the coefficient space. For instance, if $t_1$ is sampled along the unit circle and $t_2$ within the disk, the variable $t_1$ governs the angular phase of the coefficients, while $t_2$ modulates their radial magnitude. Alternatively, if $t_1$ is drawn from an annulus and $t_2$ from a line segment, the coefficients exhibit a combination of periodic radial modulation and linear gradients. This flexibility enables exploration of hybrid coefficient structures that have not been systematically investigated before.

Once the parameter space is established, each sampled pair $\theta = (t_1,t_2)$  is treated as an independent experiment. The framework evaluates all parameters, substitutes them into the symbolic coefficient expressions, and assembles the coefficient vector [$a_n,a_{n-1},…,a_0$]. Each polynomial $p(x;\theta)$ is then solved for all complex roots. Two complementary numerical solvers are provided. The default solver employs NumPy’s companion matrix eigenvalue method, which computes all roots simultaneously using LAPACK’s QR and Schur routines in double precision. This approach is highly efficient and well suited for large parameter sweeps involving millions of polynomials. For problems requiring certified accuracy or higher precision, the framework integrates MPSolve, a multi-precision polynomial solver based on the Aberth-Ehrlich simultaneous iteration. MPSolve dynamically adjusts precision and provides residual-based verification of all computed roots. NumPy offers maximum throughput for exploratory visualization, while MPSolve ensures numerical rigor for ill conditioned or high degree cases. Together they allow the system to balance efficiency and reliability. 

The grid boundaries are determined empirically to capture the central 95\% of the data while excluding numerical outliers. A small symmetric margin is added to maintain a balanced frame, and the axes are scaled to preserve a square aspect ratio. Each root contributes to the density field through histogram binning, and non-finite or out of range values are removed before aggregation.
This approach yields a normalized density map that summarizes millions of root samples with consistent spatial resolution. The resulting field provides a compact numerical representation of global root structure, ready for rendering in the visualization stage.

The visualization process converts this numerical field into a high-resolution image that reveals both the statistical and aesthetic organization of polynomial roots. The framework offers several rendering modes that translate root density into color and texture. The ``Pure Pixel" mode produces sharp and precise visualizations suitable for quantitative analysis. The ``Smooth Glow" mode uses gradual transparency to emphasize clusters of roots, creating soft gradients that highlight local density variations. The ``Smoky Bloom" mode applies additive blending to generate layered, atmospheric compositions that accentuate symmetry and motion in the data. Color processing further refines the presentation. The normalized density values are adjusted for contrast, faint regions are suppressed to reduce visual noise, and the resulting intensities are mapped onto continuous color palettes that ensure smooth tonal transitions. These choices influence the visual tone of the image without changing the underlying data.

Together, the rendering and color mapping transform abstract numerical information into visually interpretable and aesthetically engaging compositions, demonstrating how the Polynomiogram framework bridges scientific visualization and digital art.

\section{Experimental results and Discussion}
This section demonstrates the versatility of the Polynomiogram framework through three complementary perspectives: proof of correctness, educational exploration, and creative expression. First, the framework is validated against classical analytical results for well-studied polynomial families, including random polynomials following the Kac distribution and the recursive Lucas polynomials, confirming its accuracy in reproducing known root structures. Next, it is applied as an educational and experimental tool to visualize bifurcation patterns in cubic systems, illustrating how parameter variation influences stability and root organization. Finally, two examples highlight its capacity for generative and artistic visualization, showing how mathematical precision and aesthetic design can coexist within the same computational platform.

\subsection{Proof of Correctness: Random Polynomials and the Kac Distribution}
Validation of the Polynomiogram framework was conducted by reproducing the known statistical properties of random polynomial ensembles. In the Kac model \cite{kac1943average}:
\begin{align}
    P_n(z) &= \sum_{k=0}^{n} a_k z^k \\
    a_k &\sim \mathcal{N}(0, 1) \quad \text{i.i.d.}
\end{align}
Lam and Nguyen recently established that the expected number of real zeros satisfies \cite{lam2025first}:
\begin{equation}
    \mathbb{E}[N_{real}] \sim \frac{2}{\pi} \log n + C_{\xi} + o(1)
\end{equation}
where $C_{\xi}$ is a distribution dependent constant that is non universal and varies continuously with the law of $\xi$. In addition, the famous analysis by Shepp and Vanderbei provided an explicit expression for the expected zero-density function $h_n(x, y)$ in the complex plane and demonstrated that, for large $n$, the zeros concentrate in a narrow annulus near $|z| = 1$ with asymptotically uniform angular distribution \cite{shepp1995complex}. This concentration represents the limiting intensity of complex zeros, forming a nearly uniform ring on the unit circle.

\begin{figure}[H]
    \centering
    \includegraphics[width=0.8\textwidth]{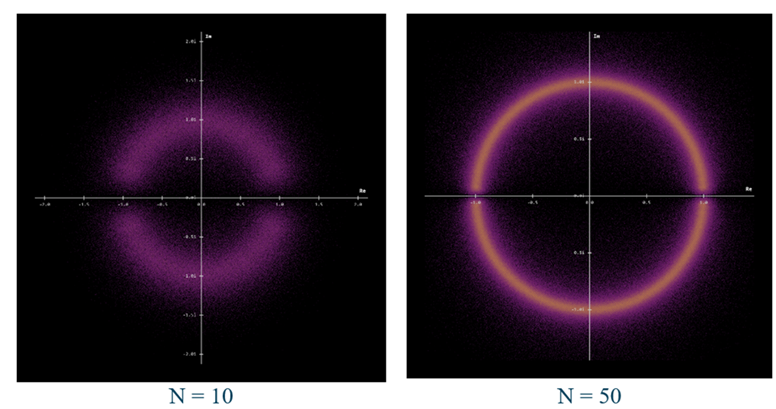} 
    \caption{Root density of random Gaussian polynomials (Kac ensemble) for degrees ($N = 10$) and ($N = 50$). The zeros converge toward the unit circle ($z=1$) as $N$ increases.}
    \label{fig:kac}
\end{figure}

The validation experiment generated $10^5$ random polynomials for each degree $n = 10$ and $n = 50$, with coefficients sampled from the standard normal distribution. All roots were computed using the companion-matrix eigenvalue method and aggregated into normalized two-dimensional density grids. The resulting Polynomiograms (Figure \ref{fig:kac}) display a bright circular band centered at the origin with radius $z \approx 1$. The interior region $z < 0.8$ remains largely vacant, and a faint horizontal trace along the real axis corresponds to the small number of real zeros predicted by Phuc’s logarithmic law. The angular distribution of zeros is uniform, and the radial histogram peaks at $r = 1.00 \pm 0.05$. For $n = 10$, the band appears diffuse due to finite-degree effects, whereas for $n = 50$, the zeros form a sharply defined ring. The observed distributions are consistent with the limiting zero measure supported on the unit circle predicted by Shepp and Vanderbei and with Kac’s theoretical expectations for the real-axis statistics. The results confirm that the Polynomiogram framework accurately reproduces both the real and complex asymptotic behavior of random polynomial ensembles and provides a reliable computational means for visualizing their root density structures.

\subsection{Validation Using Lucas Polynomials}
A second validation used a deterministic polynomial family with a known analytic zero distribution. The Lucas polynomials were selected for this purpose because they possess a closed analytical form for both the polynomial and its zeros \cite{hoggatt1973roots}. This benchmark ensured that the solver accurately reproduced an exact deterministic distribution.

The Lucas polynomials are defined by the recurrence:
\begin{equation}
    L_0(x) = 2, \quad L_1(x) = x, \quad L_{n+1}(x) = x L_n(x) + L_{n-1}(x), \quad n \ge 1
\end{equation}
A closed form can be written as:
\begin{equation}
    L_n(x) = \left(\frac{x + \sqrt{x^2+4}}{2}\right)^n + \left(\frac{x - \sqrt{x^2+4}}{2}\right)^n
\end{equation}
All zeros of $L_n(x)$ are given exactly by:
\begin{equation}
    x_k = 2i \cos\left(\frac{(2k+1)\pi}{2n}\right), \quad k=0, 1, \dots, n-1
\end{equation}
Each zero lies on the imaginary axis between $(-2i)$ and $(2i)$ and the zeros are symmetric with respect to the origin. As $n \to \infty$, the zeros become dense on the segment connecting these two points.

\begin{figure}[H]
    \centering
    \includegraphics[width=0.6\textwidth]{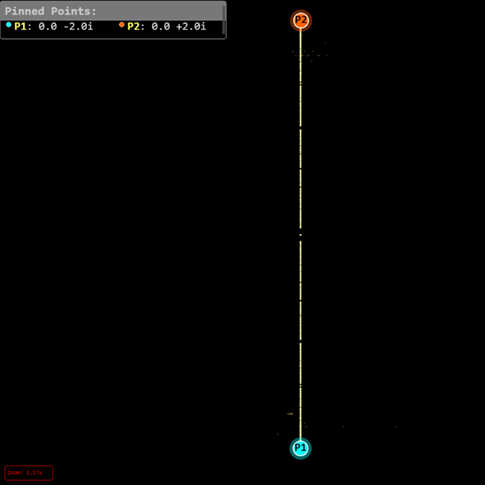} 
    \caption{Numerical root distribution of the Lucas polynomials for degrees up to $N = 1000$ computed with MP Solve at 500-digit precision.}
    \label{fig:lucas}
\end{figure}

The numerical visualization produced by the application exhibited a vertical alignment of zeros centered on the imaginary axis, confirming that all computed roots satisfy $\text{Re}(x) \approx 0$ within numerical precision (see Figure \ref{fig:lucas}). The upper and lower extremities of the distribution were located near $(\pm 2i)$, consistent with the theoretical bounds of the Lucas polynomial zeros. The computed positions of selected zeros, such as $(x \approx 0 + 2.0i)$ and $(x \approx 0 - 2.0i)$, correspond closely to the analytical limits $(\pm 2i)$. The high precision results obtained using MP Solve therefore validate that the implemented solver reproduces both the geometric confinement and the statistical density predicted by theory for the Lucas polynomial sequence.

Computing the zeros of high degree Lucas polynomials require more than standard double
precision arithmetic since the zeros form a narrow vertical distribution centered on the imaginary axis. This clustering makes the eigenvalue problem highly sensitive to rounding. Conventional libraries such as NumPy rely on fixed precision linear algebra and therefore cannot maintain the symmetry, confinement, or correct extremal behavior of the zeros once the degree becomes large. MPSolve avoids these issues by using multi-precision arithmetic, certified refinement, and numerically stable algorithms for clustered roots.

\subsection{Visualization of Bifurcation and Catastrophe Structures}
In bifurcation theory, qualitative changes in the roots of an equation $P(x; \theta) = 0$ occur when the control parameters $\theta$ cross the discriminant set $\Sigma = \{ \theta : \text{Disc}_x P(\cdot ; \theta) = 0 \}$. The discriminant identifies parameter values for which two or more roots coincide, defining the manifold that separates regions with different topological structures of the root set. The Polynomiogram supplements this parametric description by integrating over a bounded parameter domain and projecting the total solution manifold $S = \{ (\theta, z) \in \mathbb{R}^p \times \mathbb{C} : P(z; \theta) = 0 \}$ onto the complex plane. This projection does not follow individual root trajectories as parameters vary but instead aggregates all roots in spectral space, transforming the discriminant surface in parameter space into visible geometric structures such as ridges or discontinuities in root density. Through this representation, parametric bifurcations appear as geometric transitions in the spectral domain.

\begin{figure}[H]
    \centering
    \includegraphics[width=0.7\textwidth]{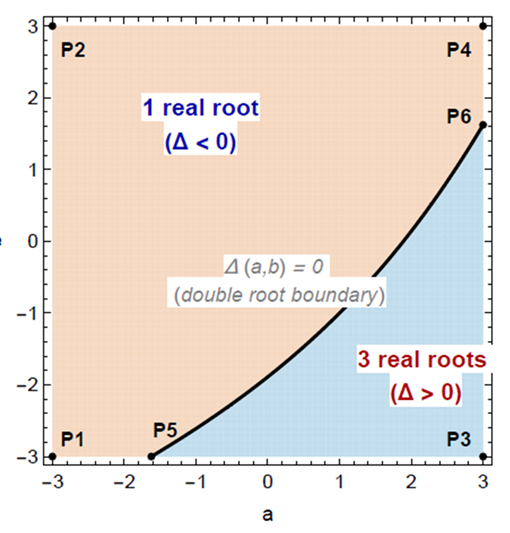} 
    \caption{Bifurcation diagram for the cubic polynomial $P(x) = x^3 + a x^2 + b x - 1$. Points $(P_1)$ through $(P_6)$ mark the intersections of the discriminant boundary $\Delta(a,b)=0$ with the limits of the parameter domain $a,b \in [-3,3]$. The black curve connecting $(P_5 - P_6)$ separates the one-real-root region ($\Delta < 0$, orange) from the three-real-root region ($\Delta > 0$, blue).}
    \label{fig:bifurcation}
\end{figure}

To illustrate the relationship, consider the cubic polynomial family:
\begin{equation}
    P(x) = x^3 + ax^2 + bx - 1
\end{equation}
with parameters $(a,b) \in [-3,3]^2$. This cubic model provides the simplest analytic case in which bifurcation structure can be computed exactly and thus serves as a reference for interpreting spectral features in the Polynomiogram.
The discriminant of this cubic is:
\begin{equation}
    \Delta(a,b) = a^2 b^2 - 4b^3 + 4a^3 - 27 - 18ab
\end{equation}
and the boundary between distinct root regimes is given by the condition $\Delta(a,b) = 0$. On this curve, the polynomial has a double real root $r$ that satisfies both $P(r) = 0$ and $P'(r) = 0$. Eliminating $a$ and $b$ yields the parametric representation:
\begin{equation}
    a(r) = -2r - \frac{1}{r^2}, \quad b(r) = r^2 + \frac{2}{r}, \quad r \neq 0.
\end{equation}
The sign of the discriminant determines the number and type of real roots:
\begin{itemize}
    \item $\Delta(a,b) > 0 \Rightarrow 3$ real roots
    \item $\Delta(a,b) < 0 \Rightarrow 1$ real root and one complex conjugate pair.
\end{itemize}
The curve $(a(r), b(r))$ separates the parameter space into these two regimes as shown in Figure \ref{fig:bifurcation}. The region above the curve, where $\Delta < 0$, corresponds to the single real root regime, and the region below, where $\Delta > 0$, corresponds to the three real root regimes. Crossing this boundary changes the number of real roots by two, representing a fold bifurcation that separates monostable and tri-stable configurations.

To characterize the bifurcation structure of the cubic family $P(x) = x^3 + ax^2 + bx - 1$, the boundary points labeled $(P_1)$ through $(P_6)$ in Figure \ref{fig:bifurcation} represent the intersections of the discriminant curve $\Delta(a,b) = 0$ with the boundaries of the parameter domain $a,b \in [-3,3]$. Along this discriminant, the cubic possesses a double root that marks the transition between the one real root and three real roots regimes. At each labeled point $P_i$, the corresponding parameter coordinates $a,b$ and their associated roots $x$ were computed explicitly. The numerical results are summarized in Table \ref{tab:roots}, which list the parameters $a,b$ together with all distinct roots $x_A, x_B, x_C$ for the same parameter pair, where subscripts $(A,B,C)$ denote separate real or complex solutions of the cubic.

\begin{table}[H]
\centering
\caption{Representative root values in distinct parameter regimes.}
\label{tab:roots}
\begin{tabular}{@{}|c|c|c|c|l|@{}}
\multicolumn{5}{c}{\textbf{One root regime}} \\
\hline
\textbf{ID} & \textbf{Point} & \textbf{a} & \textbf{b} & \textbf{Root} $x$\\
\hline
P1A & P1 & -3.00 & -3.00 & -0.42 - 0.28*I \\
\hline
P1B & P1 & -3.00 & -3.00 & -0.42 + 0.28*I \\
\hline
P1C & P1 & -3.00 & -3.00 & 3.85 \\
\hline
P4A & P4 & 3.00 & 3.00 & -1.63 - 1.09*I \\
\hline
P4B & P4 & 3.00 & 3.00 & -1.63 + 1.09*I \\
\hline
P4C & P4 & 3.00 & 3.00 & 0.26 \\
\hline
\multicolumn{5}{c}{\textbf{Three real roots regime}} \\
\hline
\textbf{ID} & \textbf{Point} & \textbf{a} & \textbf{b} & \textbf{Root} $x$\\
\hline
P3A & P3 & 3.00 & -3.00 & -3.73 \\
\hline
P3B & P3 & 3.00 & -3.00 & -0.27 \\
\hline
P3C & P3 & 3.00 & -3.00 & 1.00 \\
\hline
P5A & P5 & -1.62 & -3.00 & -0.60 \\
\hline
P5B & P5 & -1.62 & -3.00 & 2.81 \\
\hline
P6A & P6 & 3.00 & 1.62 & -1.68 \\
\hline
P6B & P6 & 3.00 & 1.62 & 0.36 \\ 
\hline
\end{tabular}
\end{table}

Among these, $P_2 = (-3, 3)$ is of particular importance: it corresponds to the triple-root configuration $P(x) = (x-1)^3$, where the discriminant and its derivative both vanish simultaneously. This point defines the cusp of the discriminant surface and represents the unique degenerate case in which the three roots coalesce into one. In the spectral domain, the projection of $P_2$ is associated with a local drop in root density, reflecting the fact that the triple-root condition occupies zero measure in parameter space and therefore contributes negligibly to the aggregated density map. The remaining points $P_1, P_4$ lie in the one-real-root regime $\Delta < 0$ and each exhibit one real root together with a complex-conjugate pair, as shown in Table \ref{tab:roots}. Points $P_3, P_5, P_6$ lie in the three-real-roots regime $\Delta > 0$ and display three distinct real solutions. Each labeled point $P_i$ projects to a specific feature or discontinuity in the complex-root density map, enabling direct comparison between the bifurcation structure of the cubic system and the corresponding patterns visualized in the Polynomiogram.

\begin{figure}[H]
    \centering
    \includegraphics[width=0.9\textwidth]{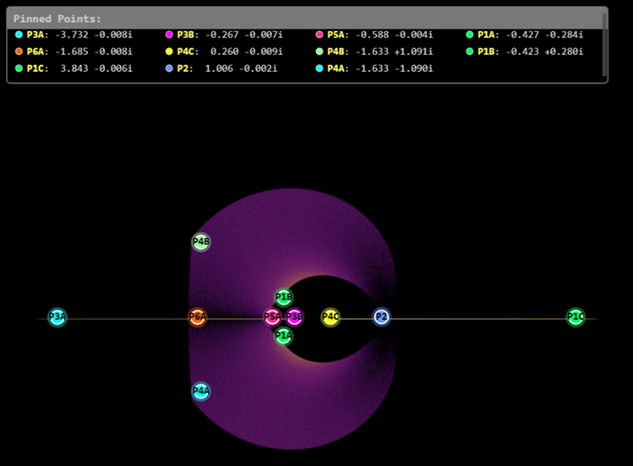} 
    \caption{Polynomiogram of the cubic polynomial $P(x) = x^3 + ax^2 + bx - 1$ showing the distribution of all roots over the parameter domain $a,b \in [-3,3]$.}
    \label{fig:cubic_poly}
\end{figure}

In Figure \ref{fig:cubic_poly}, the Polynomiogram of the cubic family $P(x) = x^3 + ax^2 + bx - 1$, the structure of the real and imaginary projections is constrained by the boundary points $P_1 - P_6$. Along the real axis the support extends from $P_{3A} \approx -3.732$ to $P_{1C} \approx 3.843$, which are the extreme real roots attainable within the parameter box $a,b \in [-3,3]$. The real-axis density is not continuous: there is a forbidden interval between $P_{3B} \approx -0.268$ and $P_{4C} \approx 0.260$, where roots cannot occur because the feasibility line $b = \frac{1-x^3}{x} - xa$ lies outside $[-3,3]$. The point $P_{6A} \approx -1.677$ marks the real location where the system crosses the discriminant; to the left the cubic has three real roots, while to the right it has one real root and a complex conjugate pair. The bounds on the imaginary part are read from the vertical markers: $P_{5A} \approx -0.589i$ is the inner left touch of the void, $P_{1A}$ and $P_{1B}$ form the lower and upper limits at $a=b=-3$, $P_{4B}$ and $P_{4A}$ form the corresponding limits at $a=b=3$, and the triple root $P_2 = (x-1)^3$ at $x=1$ sits at the center of the inner contour where density drops. Taken together, these labeled points specify the maximal real span, the gap on the real axis, the inner and outer bounds for the imaginary parts, and the exact abscissa of the transition from three real roots to one real root with a complex pair, so they provide a direct geometric link between the parameter-space bifurcation map and the spectral features in the root-density image.

\subsection{Algorithmic art generator}
\begin{figure}[H]
    \centering
    \includegraphics[width=1.0\textwidth]{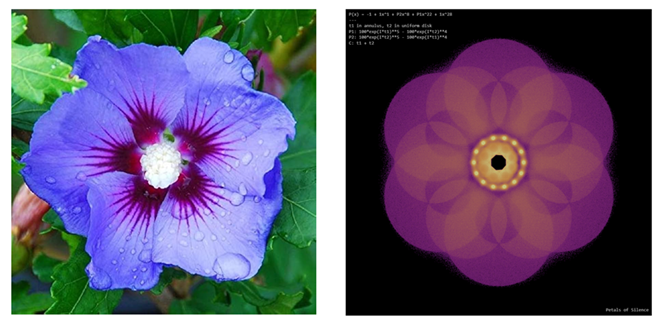} 
    \caption{Comparison between natural and algorithmically generated forms. The left panel shows a photograph of a hibiscus flower, while the right panel presents a Polynomiogram rendering produced entirely from the distribution of polynomial roots, \textit{Petals of Silence}.}
    \label{fig:hibiscus}
\end{figure}

An illustrative example is shown in Figure \ref{fig:hibiscus}, featuring the Polynomiogram artwork \textit{Petals of Silence}, in which the platform is used to reproduce the form of a hibiscus flower. The image on the left displays a photograph of the real flower, while the image on the right shows the rendering generated from the polynomial:
\begin{equation}
    P(x) = -1 + x + P_2 x^8 + P_1 x^{22} + x^{28}
\end{equation}
The coefficients are defined as:
\begin{equation}
    P_1 = 100 e^{i 5 t_1} - 100 e^{i 4 t_2}, \quad P_2 = 100 e^{i 5 t_2} - 100 e^{i 4 t_1}
\end{equation}
with $t_1$ sampled from an annulus and $t_2$ sampled from a uniform disk in the complex plane. These heterogeneous sampling domains enforce distinct magnitude phase distributions for each coefficient, thereby introducing asymmetry and interference patterns that drive the multi-scale, petal like geometry of the rendering. The correspondence of the computational form with the natural flower morphology demonstrates how coefficient domain engineering in the root-distribution framework can translate algebraic structure into visual features.

\begin{figure}[H]
    \centering
    \includegraphics[width=0.8\textwidth]{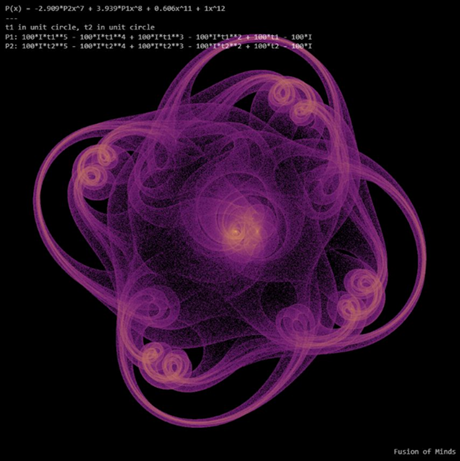} 
    \caption{Polynomiogram artwork, \textit{Fusion of Minds}. The interleaved phase terms induce constructive and destructive interference between coefficient magnitudes, yielding a multi-scale, rotationally entangled structure.}
    \label{fig:fusion}
\end{figure}

Beyond natural forms, the Polynomiogram framework can also encode conceptual or experiential themes through its parameterization. By selecting coefficient domains that represent specific relationships or symbolic structures, the system enables the generation of personalized and thematic compositions. Figure \ref{fig:fusion} illustrates this capability through the artwork \textit{Fusion of Minds}, which pays tribute to the sequence of innovations that revolutionized artificial intelligence and modern language modeling. The composition expresses gratitude for the collective progress that has transformed both science and society.

The piece was designed to represent the intellectual convergence of four landmark contributions in neural language processing. \textit{Word2Vec} introduced in \cite{mikolov2013efficient} distributed vector representations that captured semantic meaning. \textit{Seq2Seq} developed the encoder and decoder framework that enabled sequence translation \cite{sutskever2014sequence}. \textit{Transformer} presented the attention mechanism that made large-scale model training practical \cite{vaswani2017attention}. \textit{BERT} demonstrated how pretraining and fine-tuning could generalize language models across many tasks \cite{devlin2019bert}.

The composition symbolizes the convergence of independent ideas into a unified field, reflecting the collaborative synthesis that drives the evolution of modern language models. To express appreciation for these contributions, a polynomial was defined using \textit{Word2Vec} as the reference for coefficient scaling. The degree of each term was set to \textbf{($2025$ -  publication year)}, and the coefficient magnitude was given by the ratio of the paper’s citation count to that of \textit{Word2Vec}. This produces the normalized polynomial:
\begin{equation}
    p(x) = 1.000 x^{12} + 0.606 x^{11} + 3.939 x^8 + 2.909 x^7
\end{equation}
Rendering this polynomial with the Polynomiogram created the artwork \textit{Fusion of Minds}, shown in Figure \ref{fig:fusion}. The visualization symbolizes how independent ideas from different researchers and countries merge into a collective center of innovation. Each curved path represents a single scientific milestone, and their convergence illustrates the cumulative nature of discovery. Within this composition, the interwoven coefficient phases symbolize the interaction of independent ideas that collectively shaped the current paradigm of large-scale language understanding.

\section{Conclusion}
This work presented a unified framework for computing, visualizing, and interpreting polynomial roots through the Polynomiogram platform. The approach combines symbolic parameterization, efficient numerical root solving, and density-based visualization into a single computational environment. Verification using the Kac and Lucas ensembles confirmed the accuracy of the numerical implementation, reproducing known statistical and analytic results. The bifurcation analysis of the cubic polynomial $P(x) = x^3 + ax^2 + bx - 1$ demonstrated how discriminant boundaries in parameter space correspond to distinct geometric and spectral features in the complex plane. By mapping the parameter intersections $P_1-P_6$ and their projected roots, the connection between classical bifurcation theory and the spectral representation of root distributions was established.

Beyond analysis, the Polynomiogram was shown to be an adaptable platform for artistic generation. These results highlight the versatility of the framework as both a scientific visualization tool and a medium for creative design. Future research will extend the polynomiogram's underlying soft computational model into higher-dimensional parameter spaces, moving beyond the current complex plane representation. A key technical focus involves the integration of adaptive sampling methodologies, leveraging techniques from stochastic optimization or Fuzzy Control Systems to ensure computational efficiency and precision, particularly in regions exhibiting complex fractal-like behavior or high sensitivity to initial conditions. Furthermore, the rendering engine will be significantly augmented to facilitate dynamic visualization through real-time animation support and interactive explorability.

\section*{Code Availability}

The complete implementation of the Polynomiogram platform, including data generation scripts and visualization examples, is publicly available at: https://github.com/gibbenergy/Polynomial-Root-Art-Generator




\bibliographystyle{elsarticle-num-names} 
\bibliography{references}

@book{briggs1992fractals,
  title={Fractals: The patterns of chaos: A new aesthetic of art, science, and nature},
  author={Briggs, John},
  year={1992},
  publisher={Simon and Schuster}
}

@book{kuznetsov1998elements,
  title={Elements of applied bifurcation theory},
  author={Kuznetsov, Yuri A},
  year={1998},
  publisher={Springer}
}

@article{schinzel1970reducibility,
  title={Reducibility of lacunary polynomials II},
  author={Schinzel, Andrzej},
  journal={Acta Arithmetica},
  volume={16},
  number={4},
  pages={371--392},
  year={1970}
}

@article{nalli2009generalized,
  title={On generalized Fibonacci and Lucas polynomials},
  author={Nalli, Ayse and Haukkanen, Pentti},
  journal={Chaos, Solitons \& Fractals},
  volume={42},
  number={5},
  pages={3179--3186},
  year={2009}
}

@article{moore1959interval,
  title={Interval analysis I},
  author={Moore, Ramon E and Yang, CT},
  journal={Technical Document LMSD-285875, Lockheed Missiles and Space Division, Sunnyvale, CA, USA},
  year={1959}
}

@article{kac1943average,
  title={On the average number of real roots of a random algebraic equation},
  author={Kac, Mark},
  journal={Bull. Amer. Math. Soc.},
  volume={49},
  number={4},
  pages={314--320},
  year={1943}
}

@article{babai1979monte,
  title={Monte-Carlo algorithms in graph isomorphism testing},
  author={Babai, L{\'a}szl{\'o}},
  journal={Universit{\'e} tde Montr{\'e}al Technical Report, DMS},
  number={79-10},
  year={1979}
}

@book{koshy2019fibonacci,
  title={Fibonacci and Lucas Numbers with Applications, Volume 2},
  author={Koshy, Thomas},
  volume={2},
  year={2019},
  publisher={John Wiley \& Sons}
}

@book{peitgen1986beauty,
  title={The beauty of fractals: images of complex dynamical systems},
  author={Peitgen, Heinz-Otto and Richter, Peter H},
  year={1986},
  publisher={Springer Science \& Business Media}
}

@book{mason2002chebyshev,
  title={Chebyshev polynomials},
  author={Mason, J.C. and Handscomb, D.C},
  year={2022},
  publisher={Chapman and Hall/CRC}
}

@inproceedings{kaltofen1995subquadratic,
  title={Subquadratic-time factoring of polynomials over finite fields},
  author={Kaltofen, Erich and Shoup, Victor},
  booktitle={Proceedings of the twenty-seventh annual ACM symposium on Theory of computing},
  pages={398--406},
  year={1995}
}

@article{kalantari2005polynomiography,
  title={Polynomiography: from the fundamental theorem of Algebra to art},
  author={Kalantari, Bahman},
  journal={Leonardo},
  volume={38},
  number={3},
  pages={233--238},
  year={2005}
}

@book{trefethen2022numerical,
  title={Numerical linear algebra (25th edition)},
  author={Trefethen, L.N. and Bau, D.},
  year={2022},
  publisher={SIAM}
}

@article{edelman1995many,
  title={How many zeros of a random polynomial are real?},
  author={Edelman, Alan and Kostlan, Eric},
  journal={Bulletin of the American Mathematical Society},
  volume={32},
  number={1},
  pages={1--37},
  year={1995}
}

@book{redei2014lacunary,
  title={Lacunary polynomials over finite fields},
  author={R{\'e}dei, L{\'a}szl{\'o}},
  year={2014},
  publisher={Elsevier}
}

@article{edelman1995polynomial,
  title={Polynomial roots from companion matrix eigenvalues},
  author={Edelman, Alan and Murakami, Hiroshi},
  journal={Mathematics of Computation},
  volume={64},
  number={210},
  pages={763--776},
  year={1995}
}

@book{press2007numerical,
  title={Numerical recipes 3rd edition: The art of scientific computing},
  author={Press, William and others},
  year={2007},
  publisher={Cambridge university press}
}

@article{good1961colleague,
  title={The colleague matrix, a Chebyshev analogue of the companion matrix},
  author={Good, IJ},
  journal={The Quarterly Journal of Mathematics},
  volume={12},
  number={1},
  pages={61--68},
  year={1961}
}

@article{fiedler2003note,
  title={A note on companion matrices},
  author={Fiedler, Miroslav},
  journal={Linear Algebra and its Applications},
  volume={372},
  pages={325--331},
  year={2003}
}

@inproceedings{mikolov2013efficient,
  title={Efficient estimation of word representations in vector space},
  author={Mikolov, Tomas and others},
  booktitle={ICLR 2013 (Workshop Poster)},
  year={2013}
}

@article{sutskever2014sequence,
  title={Sequence to sequence learning with neural networks},
  author={Sutskever, Ilya and Vinyals, Oriol and Le, Quoc V},
  journal={Advances in neural information processing systems},
  volume={27},
  year={2014}
}

@article{lam2025first,
  title={On the First Non-Universal Term in Random Polynomial Real Zeros},
  author={Lam, Phuc and Nguyen, Oanh},
  journal={arXiv preprint arXiv:2509.12170},
  year={2025}
}

@article{shepp1995complex,
  title={The complex zeros of random polynomials},
  author={Shepp, Larry A and Vanderbei, Robert J},
  journal={Transactions of the American Mathematical Society},
  volume={347},
  number={11},
  pages={4365--4384},
  year={1995}
}

@article{vaswani2017attention,
  title={Attention is all you need},
  author={Vaswani, A and others},
  journal={Advances in neural information processing systems},
  volume={30},
  year={2017}
}

@article{hoggatt1973roots,
  title={Roots of Fibonacci polynomials},
  author={Hoggatt Jr, Verner E and Bicknell, Marjorie},
  journal={The Fibonacci Quarterly},
  volume={11},
  number={3},
  pages={271--274},
  year={1973},
  publisher={Taylor \& Francis}
}

@inproceedings{devlin2019bert,
  title={Bert: Pre-training of deep bidirectional transformers for language understanding},
  author={Devlin, Jacob and others},
  booktitle={Proceedings of the 2019 conference of the North American chapter of the association for computational linguistics (NAACL 2019), Vol. 1},
  pages={4171--4186},
  year={2019}
}

@article{jenkins1970three,
  title={A three-stage algorithm for real polynomials using quadratic iteration},
  author={Jenkins, Michael A and Traub, Joseph F},
  journal={SIAM Journal on Numerical Analysis},
  volume={7},
  number={4},
  pages={545--566},
  year={1970},
  publisher={SIAM}
}

@misc{henrlci1977applied,
  title={Applied and computational complex analysis},
  author={Henrlci, P},
  year={1977},
  publisher={New York, John Wiley \& Sons}
}

@article{moniruzzaman2025forced,
  title={Forced dynamic operation of propylene selective oxidation to acrolein on bismuth-molybdate structured catalysts},
  author={Moniruzzaman, Mohammad and Grabow, Lars C and Harold, Michael P},
  journal={Applied Catalysis A: General},
  volume={691},
  pages={120034},
  year={2025},
  publisher={Elsevier}
}

@article{sun2018bifurcation,
  title={Bifurcation analysis of methane oxidative coupling without catalyst},
  author={Sun, Zhe and Kota, Arun and Sarsani, Sagar and others},
  journal={Chemical Engineering Journal},
  volume={343},
  pages={770--788},
  year={2018},
  publisher={Elsevier}
}

@inproceedings{collins1976polynomial,
  title={Polynomial real root isolation using Descarte's rule of signs},
  author={Collins, George E and Akritas, Alkiviadis G},
  booktitle={Proceedings of the third ACM symposium on Symbolic and algebraic computation},
  pages={272--275},
  year={1976}
}



\end{document}